\documentclass[prd, twocolumn, nofootinbib, floatfix]{revtex4}

\usepackage{epsfig} 
\usepackage{amsmath}
\usepackage{amsbsy,color}

\newcommand{\beq}{\begin{equation}}
\newcommand{\eeq}{\end{equation}}
\newcommand{\beqa}{\begin{eqnarray}}
\newcommand{\eeqa}{\end{eqnarray}}

\newcommand{\cvis}{{c_{\rm vis}}}
\newcommand{\ceff}{{c_{\rm eff}}}
\newcommand{\cvisSQ}{{c^{2}_{\rm vis}}}
\newcommand{\ceffSQ}{{c^{2}_{\rm eff}}}
\newcommand{\neff}{{N_{\rm eff}}}

\begin{document} 

\title{Constraints on neutrino and dark radiation interactions using cosmological observations} 
\author{Tristan L.\ Smith, Sudeep Das and  Oliver Zahn} 
\affiliation{Berkeley Center for Cosmological Physics \& Berkeley Lab, 
University of California, Berkeley, CA 94720, USA} 
\date{\today}

\begin{abstract} 
Observations of the cosmic microwave background (CMB) and large-scale 
structure (LSS) provide a 
unique opportunity to explore the fundamental properties of the constituents that compose the 
cosmic dark radiation background (CDRB), of which the three standard neutrinos are thought to be the dominant component. 
We report on the first constraint to the CDRB
rest-frame sound speed, $\ceffSQ$, using the most recent CMB and LSS 
data. Additionally, we report improved constraints to the CDRB viscosity 
parameter, $\cvisSQ$.   For a non-interacting species, these parameters both equal 1/3.  Using current data we find that 
a standard CDRB, composed entirely of three non-interacting neutrino species, is ruled out at the $99\%$ confidence 
level (C.L.) with $\ceffSQ = 0.30_{-0.026}^{+0.027}$ and $\cvisSQ =0.44^{+0.27}_{-0.21}$ (95\% C.L.).  
We also discuss how constraints to these parameters from 
current and future observations (such as the Planck satellite) allow us to explore the 
fundamental properties of any anomalous 
radiative energy density beyond the standard three neutrinos. 
\end{abstract} 

\maketitle 

\section{Introduction} 

A complete understanding of the basic building blocks of the universe 
hinges on understanding the elusive properties of neutrinos.  As a result 
of having extremely 
weak interactions, neutrinos are 
the least accurately measured of the known fundamental particles. However, even with 
our limited knowledge of neutrino properties, the fact that they are massive \cite{mass}
represents one of the most significant challenges to the 
Standard Model of particle physics.  Therefore, any improved knowledge of 
the properties of neutrinos will not only serve to shed new light on a relatively poorly explored 
aspect of fundamental physics but may also provide further evidence of inadequacies of the Standard Model. 
Adding urgency to the exploration of the properties of neutrinos, a recent combination of data on anomalous 
neutrino mixing \cite{mixing} along with observations of the neutrino 
flux from nuclear reactors \cite{reactor}
has indicated anomalous mixing between neutrino flavors.  

An important consequence  
of these modifications to the neutrino sector is their imprint on cosmological observations.  
Within the Standard Model, a cosmological 
background of neutrinos is in thermodynamic equilibrium with the other cosmological species for temperatures $T_{\nu} \gtrsim 1$ MeV, after which the neutrino background decouples and only interacts 
through gravity. 
Neutrinos are thought to comprise a significant fraction of the radiative energy density during big bang nucleosynthesis (BBN) causing a 
measurable effect on the abundances of 
the primordial light elements  \cite{BBN} as well as during 
the formation of the cosmic microwave background (CMB) and large scale structure (LSS) \cite{perturb}.  

Cosmological observations are able to place precise constraints on the effective number of neutrino species, $\neff$, 
defined 
so that the total radiative energy density is given by $\rho_{\rm rad} = \rho_{\gamma}[1+
\neff (7/8)(4/11)^{4/3}]$, where $\rho_{\gamma}$ is the energy density in photons.  The cosmological radiative 
content in addition to the photons is known as the cosmic dark radiation background (CDRB). 
In the standard cosmological model the only 
radiative energy density besides photons are the three known neutrino species so that $\neff = 3.046$, with
the small correction due to finite temperature QED effects
and neutrino flavor mixing \cite{QED}.  

Greatly adding to the intrigue, a combination of the most current observations of the CMB and LSS  
indicate that $\neff >3$ at the 99.9\% C.L., with $\neff = 4.0^{+0.58}_{-0.57}$ (95\% C.L., see Table I).  
Cosmological constraints on $\neff$ 
are insensitive to anything but the total energy density contained in the CDRB.   Most radiative 
backgrounds, including neutrinos, share the property that they are `non-interacting', i.e., they only interact with 
other cosmological species through gravity.  Examples of such backgrounds include axions \cite{Turner:1986tb} and short-wavelength gravitational-waves \cite{Smith:2006nka}.
Any further interpretation 
of finding $\neff >3$ requires information on other properties of the anomalous radiative 
energy density. 

Here we constrain the values of
two additional parameters which determine the properties of the CDRB: the rest-frame 
sound speed, $\ceff$, and a viscosity parameter, $\cvis$ \cite{Hu:1998kj}.  
As we will describe further, a standard, non-interacting, radiative background has $(\ceffSQ,\cvisSQ) = (1/3,1/3)$.  
However, if the CDRB is composed of any non-standard species with significant interactions these parameters can take on different 
values (see, e.g., Refs.~\cite{Bell:2005dr,Sawyer:2006ju}) which can have a significant impact on the observed CMB power spectrum, as shown 
in Fig.~\ref{fig:Cls}.  If both $\cvisSQ$ and $\ceffSQ$  are found to be consistent with 
their standard value of $1/3$ this would lend weight to the interpretation that observations indicate 
the existence of extra relativistic, non-interacting (i.e., neutrino-like) degrees of freedom.  On the other hand, if either is found to be 
inconsistent with their standard values, any inferred anomalous radiative energy density cannot be 
composed of standard neutrinos (and may actually be the result of unaccounted for systematic effects). 

Several previous studies have looked at observational constraints on neutrino 
interactions \cite{Hannestad:2004qu,Bell:2005dr,previousCVIS}.  In particular, Refs.~\cite{previousCVIS} have used measurements 
of the CMB and LSS to constrain the value of $\cvis$.  However, this is the first study 
to place a constraint 
on the rest-frame sound speed, 
$\ceff$.  Constraints on $\ceff$ are particularly interesting since models of neutrino interactions 
indicate it can differ from its canonical value by $\sim 30\%$ (e.g.,~Ref.~\cite{Bell:2005dr}) and \emph{current} observations 
can constrain its value to $\sim 10\%$ at the 95\% confidence level (C.L.). 
Additionally, we report significantly improved constraints  on $\cvis$ by using the most recent measurements of the CMB and 
LSS lowering the uncertainty on $\cvisSQ$ by a factor of $\sim 1.5$.  

\section{Parameterization}

 The modified evolution equations for the neutrino perturbations \footnote{Although in 
 this section we refer specifically to neutrinos, these equations apply without modification to 
 any massless cosmological component.} are \cite{Hu:1998kj}
\begin{eqnarray}
\dot{\delta}_{\nu}+k \left(q_{\nu}+\frac{2}{3k} \dot{h}\right)&=& \frac{\dot{a}}{a} (1-3 \ceffSQ) \left(\delta_{\nu}+3 \frac{\dot{a}}{a}\frac{q_{\nu}}{k}\right), \\
\dot{q}_{\nu} + \frac{\dot{a}}{a}q_{\nu}+ \frac{2}{3} k \pi_{\nu}&=& k \ceffSQ \left(\delta_{\nu}+3 \frac{\dot{a}}{a}\frac{q_{\nu}}{k}\right), \\
\dot{\pi}_{\nu} +\frac{3}{5} k F_{\nu,3} &=& 3 \cvisSQ \left(\frac{2}{5} q_{\nu} + \frac{8}{15} \sigma\right), \\
\frac{2l+1}{k}\dot{F}_{\nu,l} -l F_{\nu,l-1}&=&  - (l+1) F_{\nu,l+1},\ l \geq 3,
 \end{eqnarray}
  where the dot indicates a derivative with respect to conformal time, $a$ is the scale-factor, $k$ is the wavenumber, $\ceff$ is the rest-frame sound-speed, $\cvis$ is a viscosity parameter, $\delta_{\nu}$ is the neutrino density contrast, $q_{\nu}$ is the neutrino velocity 
perturbation, $\pi_{\nu}$ is the neutrino anisotropic stress, $F_{\nu,l}$ are higher 
order moments of the neutrino distribution function, the shear, $\sigma = 1/(2k)
(\dot{h} + 6 \dot{\eta})$, $h$ and $\eta$ are the scalar metric perturbations in 
synchronous gauge \cite{ma_bert}, and the higher order moments of the distribution 
function are truncated with appropriate boundary conditions (see, e.g., Ref.~\cite{CAMB}).  

\begin{figure}[htbp]
\resizebox{!}{7cm}{\includegraphics{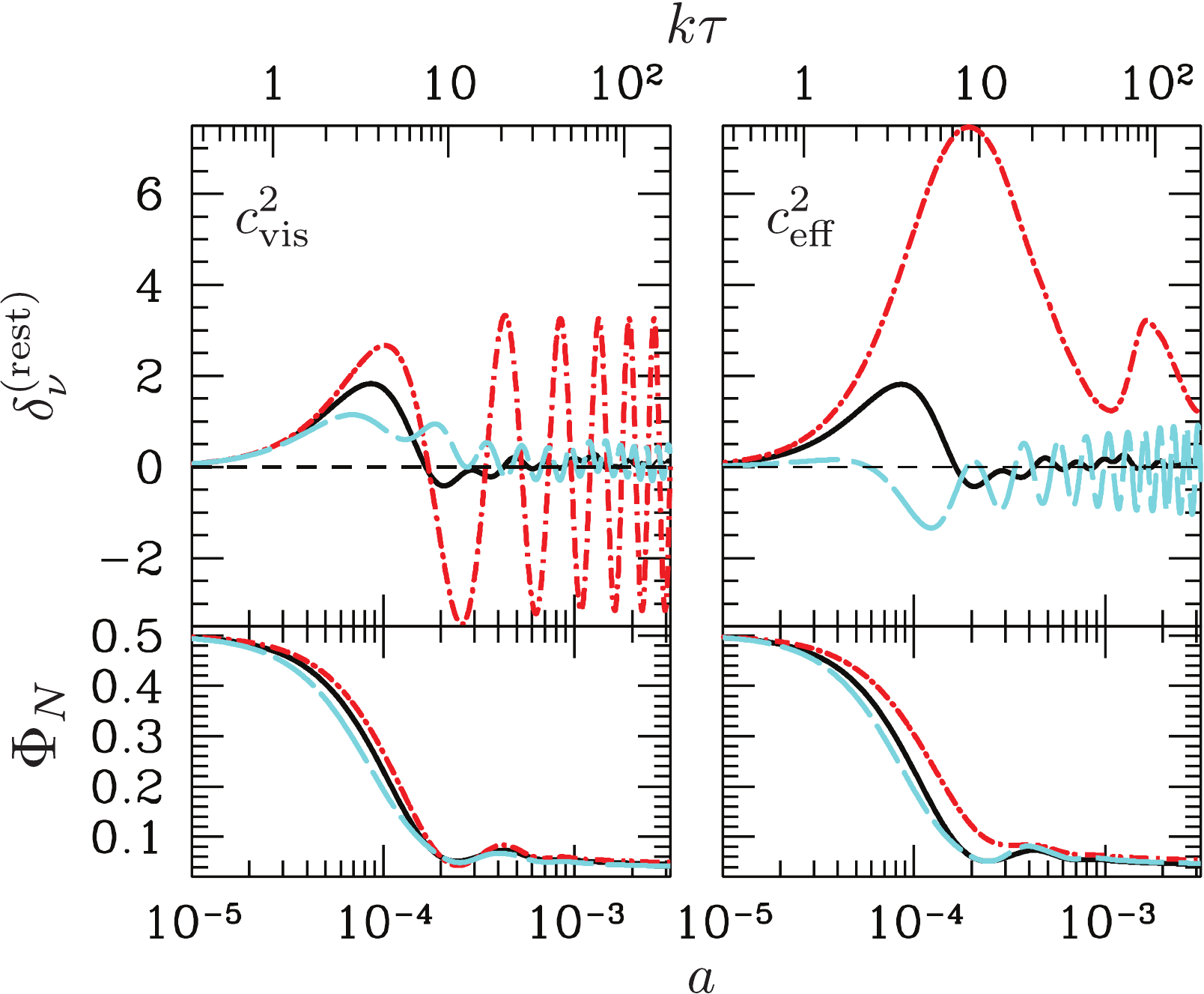}}
\caption{The evolution of the neutrino density perturbation in its rest-frame for a mode with $k=0.1\ h {\rm Mpc}^{-1}$ where $h$ is the Hubble parameter in units of 100 km/(s Mpc), as a function of the scale-factor, $a$, or 
the conformal time, $\tau$.  The black solid
curve gives the evolution for the standard case, i.e., when $\cvisSQ =\ceffSQ= 1/3$.  The 
left-hand panel shows the evolution when $\cvisSQ = 0$ (red, dot-dashed) and $\cvisSQ = 1$ (blue, dashed) with $\ceffSQ =1/3$.  With $\cvisSQ = 0$ (red, dot-dashed) the CDRB becomes a perfect fluid leading to undamped acoustic oscillations. 
The right-hand panel shows the evolution when $\ceffSQ = 0.1$ (red, dot-dashed) and $\ceffSQ = 0.8$ (blue, dashed) with $\cvisSQ =1/3$. When $\ceffSQ$ is small (red, dot-dashed) the CDRB is partially able to overcome its internal pressure support and nearly cluster.  The bottom panels show the corresponding evolution of the Newtonian potential, $\Phi_N$.}
\label{fig:evo}
\end{figure}

We note that the modified evolution equations imply a modified set of initial conditions for the 
perturbation equations since neutrinos are a significant fraction of the 
total radiative energy density at early times when the initial conditions are set. 
Following the derivation outlined in Ref.~\cite{ma_bert} we set the initial conditions to the growing mode which reverts 
to the standard adiabatic 
mode when $\cvisSQ =\ceffSQ = 1/3$ as shown in Appendix A.  

\begin{figure}[htbp]
\resizebox{!}{6cm}{\includegraphics{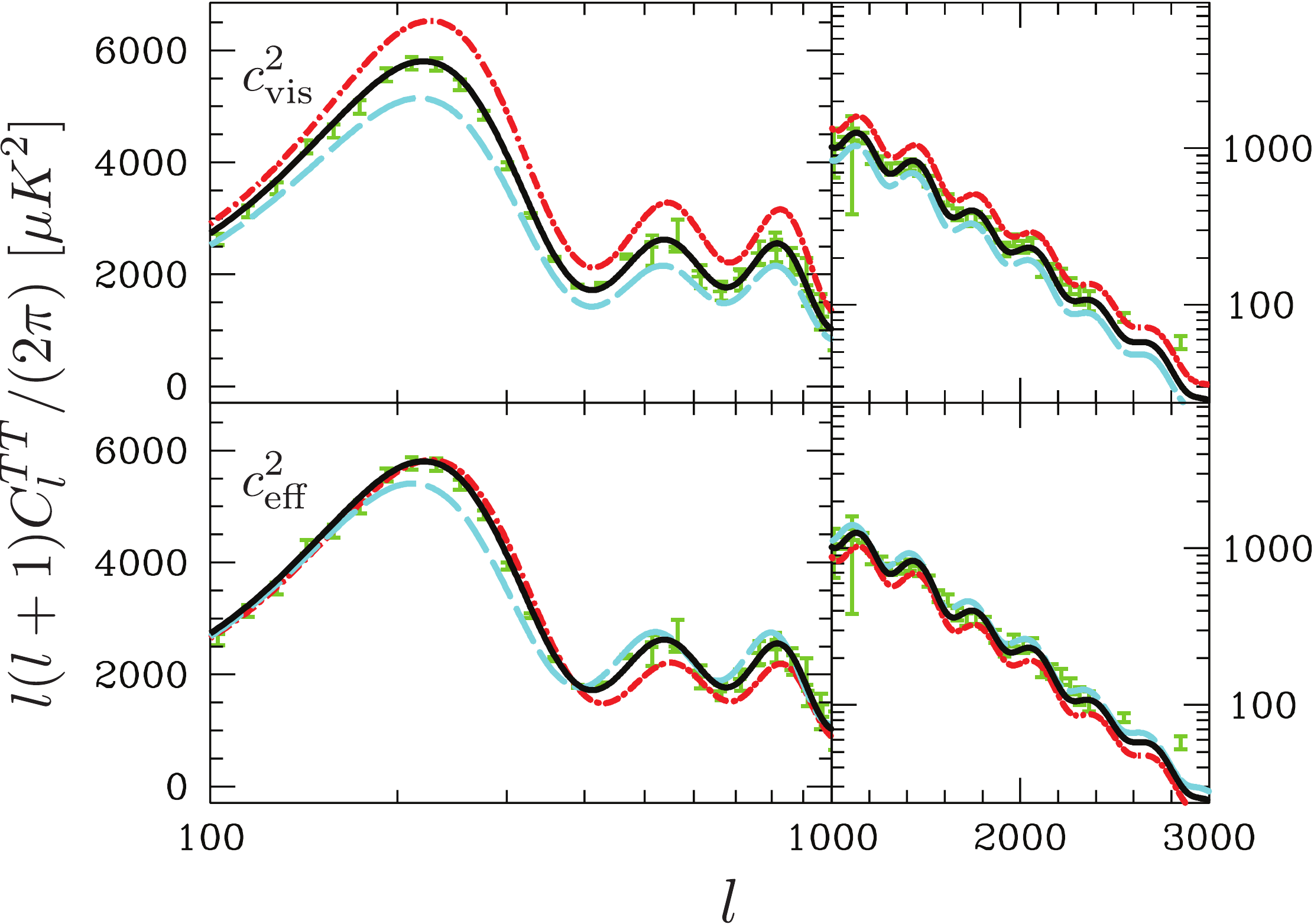}}
\caption{Modifications to the CMB temperature 
power-spectrum, $C_l^{TT}$, as both $\cvisSQ$ (top panel) and $\ceffSQ$ (bottom panel) 
are varied in the same way as in Fig.~\ref{fig:evo}: the black solid
curve gives the evolution for the standard case; the top panel shows $C_l^{TT}$ when $\cvisSQ = 0$ (red, dot-dashed) and $\cvisSQ = 1$ (blue, dashed); the bottom panel shows $\ceffSQ = 0.2$ (red, dot-dashed) and $\ceffSQ =0.7$ (blue, dashed).  The large angular scale measurements 
are from the 7-year WMAP 
release \cite{WMAP7} and on small angular scales from ACT \cite{ACT}.}
\label{fig:Cls}
\end{figure}

Varying $\cvis$ modifies the ability for neutrinos to free-stream out of a gravitational potential well  \cite{Hu:1998kj,perturb}.  
When $\cvisSQ = 0$ the CDRB becomes a perfect fluid and is capable of 
supporting undamped acoustic oscillations, shown in red (dot-dashed) on the left-hand panel of Fig.~\ref{fig:evo}.  
An increased $\cvis$ leads to an overdamping of the perturbations, shown in blue (dashed) in the left-hand panel of Fig.~\ref{fig:evo}. 

Changing $\ceff$ allows for a neutrino pressure perturbation which is non-adiabatic, i.e., 
$(\delta p - \delta \rho/3)/\bar{\rho}= (\ceffSQ - 1/3) \delta_{\nu}^{({\rm rest})}$, 
where $\delta_{\nu}^{({\rm rest})}$ is the density perturbation in a frame where the neutrino 
velocity perturbation, $q_{\nu} = 0$.  A value of $\ceffSQ< 1/3$ ($\ceffSQ> 1/3$) leads to a decreased (increased) pressure 
for the CDRB in its rest-frame, which in turn causes the amplitude of the neutrino density perturbations to 
increase (decrease), as seen on the right hand side of Fig.~\ref{fig:evo} in the red, dot-dashed (blue, dashed) curve.  

This parameterization is related to a scenario in which neutrinos have a 
significant interaction cross-section.  As an example, if neutrinos tightly couple to a perfect fluid then 
$\cvisSQ = 0$ and an analogy can be 
made between our parameterization and the tightly coupled photon-baryon fluid with a constant sound speed, 
$c_s^2$, 
related to $\ceffSQ$ through, 
$3 c_s^2 \approx \left(\ceffSQ+2/3\right)$.

We show how the CMB temperature power-spectrum is 
modified in this parameterization in Fig.~\ref{fig:Cls}.  
Note that an increase (decrease) in $\ceff$ leads to an increase (decrease) in the scales at which the neutrino 
perturbations affect the CMB.  This is due to the increase (decrease) in the the neutrino sound horizon.
These parameters 
  have a similar effect on the polarization power-spectrum, not shown here.  However, 
 since the effects of the CDRB perturbations are negligible 
 by the time large-scale structure forms, the change to the matter power-spectrum is 
 negligible \cite{Hu:1998kj}. 

\section{Results}

In order to measure these parameters we used a modified version of the publicly 
available Boltzmann code, CAMB \cite{CAMB} along with the publicly available Monte Carlo Markov chain 
code, CosmoMC \cite{cosmomc}.  We used a combination of CMB and LSS data including WMAP7 \cite{WMAP7}, ACBAR \cite{ACBAR}, ACT \cite{ACT}, SPT \cite{SPT}, the Sloan Digital Sky Survey (SDSS) DR7 LRG matter power spectrum \cite{SDSSDR7}, the SDSS small-scale matter power-spectrum measured from the Lyman-alpha forest \cite{McDonald:2004xn} and the latest determination of the Hubble parameter, $H_0$, using the 
Hubble Space Telescope \cite{Riess:2011yx}.  

In addition to various combinations of $(\neff, \cvisSQ,\ceffSQ)$ we allowed the standard six cosmological parameters, $(A_s, n_s, \tau, \theta, \Omega_b h^2, \Omega_{dm} h^2)$, to vary, where 
$A_s$ is the amplitude of the primordial power-spectrum, $n_s$ is the spectral index, $\tau$ is the 
optical depth, $\theta$ is the angular acoustic scale of the CMB, $h$ is the Hubble parameter in units of 100 ${\rm km/(s\ Mpc)}$, $\Omega_b$ is the baryon density in units of the critical  density, and $\Omega_{dm}$ is 
the dark matter density in units of the critical density.  All constraints, except where noted, force the Helium fraction, $Y_p$, to be fixed by its BBN relationship to $\Omega_b h^2$ and $\neff$ \cite{Steigman:2007xt}.  We note that this parameterization only takes into account how a change in $\neff$ causes a change in the expansion rate during BBN.  If the change in $\neff$ is due to a change in the physics of the neutrino sector the functional form $Y_p(\neff)$ may not hold.  We also consider a case where $Y_p$ is an additional free parameter as discussed below. 

\begin{figure}[htbp]
\resizebox{!}{4.9cm}{\includegraphics{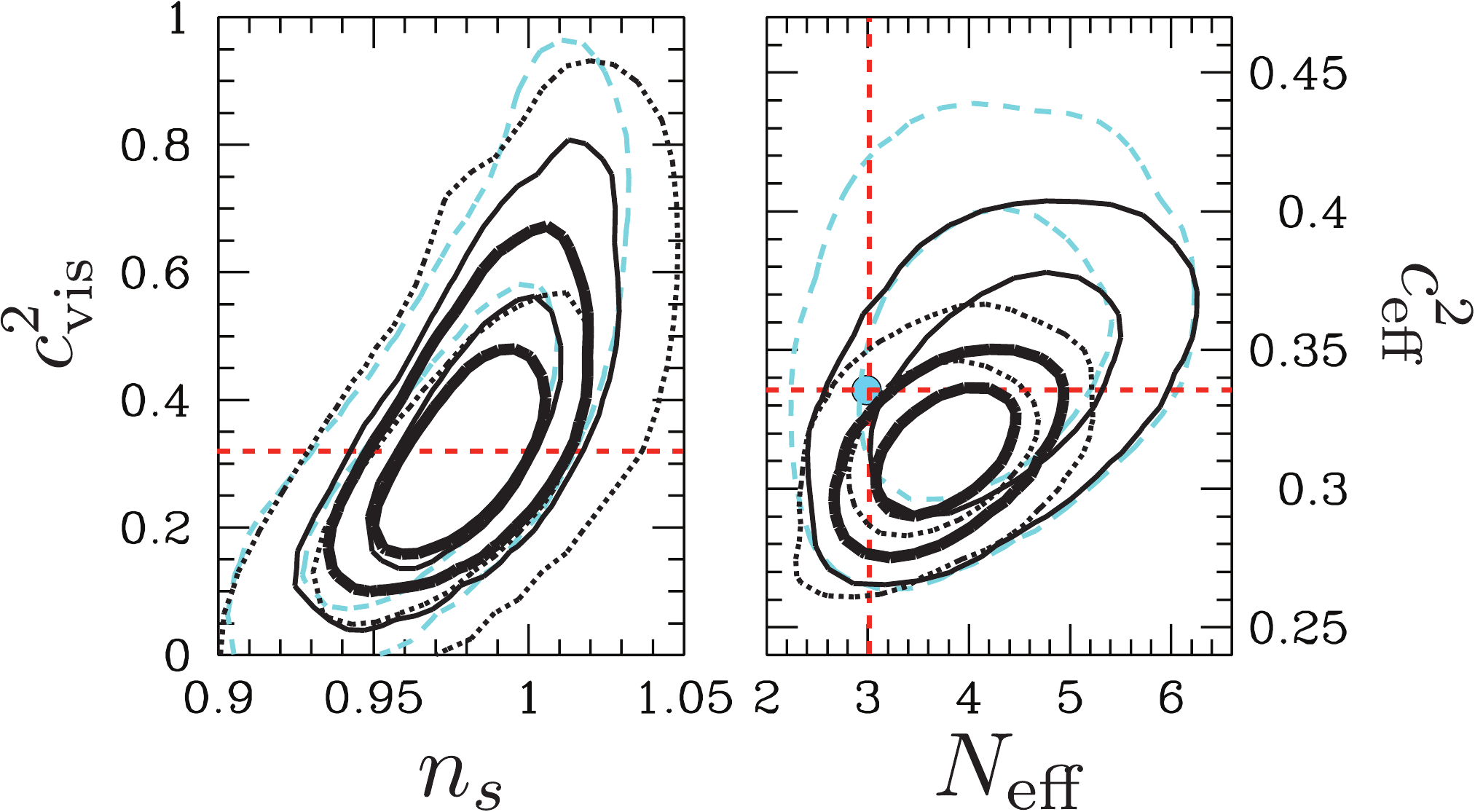}}
\caption{Two dimensional contours (68\% and 95\% C.L.) 
showing the degeneracy between 
$\cvis$/$n_s$ and $\ceff$/$\neff$.  The dotted contours show the constraints when only CMB data is used.  The blue-dashed contours show the constraints when restricting the CMB 
to just WMAP7 and large-scale structure data, excluding the Lyman-alpha data.    The thin-solid contours show constraints when restricting the CMB 
to just WMAP7 and with all large-scale structure data.  The thick contours show the constraints when 
all of the data are included.  The blue circle shows that the standard value of $\ceffSQ= 1/3$ and
$\neff = 3$ is excluded at  slightly higher than the 95\% C.L.}
\label{fig:constraints}
\end{figure}

A summary of our results can be found in Table \ref{tab:results}.   Varying the number 
of effective neutrino species, $\neff$, with $\cvisSQ=\ceffSQ = 1/3$ we recover a preference for 
an anomalous radiative energy density at the $\approx 99.9\%$ C.L.  When we allow 
 $\cvis$ and $\ceff$ to vary as well, the significance of any anomalous radiative energy 
density changes to 97.2\% C.L.  Additionally, the marginalized mean value of $\ceffSQ = 0.31 \pm 0.015$ is lower than the 
expected value of 1/3 at the  87.5\% C.L.  Therefore, as shown by the blue circle in the right-hand panel of Fig.~\ref{fig:constraints},
the standard value of $(N_{\rm eff}, \ceffSQ) = (3,1/3)$ is still disfavored at higher than the 95\% C.L.  

Since a change in $\ceff$ 
introduces a new length-scale (the neutrino sound-horizon) its affect on the CMB is only slightly correlated 
with the other parameters.  This scale-dependence is clearly shown in Fig.~\ref{fig:Cls} around the first peak.  Because of its lack of strong correlations, the observations place a 
precise constraint on $\ceffSQ$ as shown in Fig.~\ref{fig:constraints}.
Constraints to $\cvisSQ$ are not as precise because its affect on the CMB is scale-free leading to a 
degeneracy with, for example, the scalar spectral 
index, $n_s$, shown on the left-hand panel of Fig.~\ref{fig:constraints}.  

Given the tentative evidence for an anomalously large radiative energy density, it is of interest to consider the case where the number of neutrinos (with $\cvisSQ=\ceffSQ=1/3$) is fixed to three and to constrain the values of $\cvisSQ$ and $\ceffSQ$ applied to an additional one or two effective species.  We show the results of this analysis in Table I.  

\begin{table}[tb]\footnotesize
\caption{\label{tab:results}}
\begin{center}
{\sc Marginalized 1D constraints\\}
\begin{tabular}{ccccc} 
\\
\hline
\hline
 $N_{\rm eff}$ & $\cvisSQ$& $\ceffSQ$ \\ 
\hline 
$4.0^{+ 0.17+ 0.58}_{- 0.18- 0.57}$ & 1/3 &1/3\\
$3.77 ^{+ 0.18+ 0.68}_{- 0.19- 0.65}$ & $0.33^{+0.04+0.21}_{-0.06 - 0.15}$&$0.31 \pm 0.015^{+0.029}_{-0.030}$ \\
3 & $0.44 ^{+0.056+0.27}_{-0.085-0.21}$&$0.30 \pm 0.013^{+0.027}_{-0.026}$ \\
$1^{\dagger}$& $< 2.2$& $0.29\pm  0.043^{+0.1}_{-0.075}$ \\
$2^{\dagger}$& $<0.51$& $0.34 \pm  0.03 ^{+ 0.062}_{-0.052} $ \\
\hline
\end{tabular}
\end{center}
NOTES.---%
Errors are 68\%, 95\% C.L.; upper limits are 95\% C.L. \\
$^{\dagger}$ For these chains, the number of effective relativistic degrees of freedom with $\cvisSQ=
\ceffSQ=1/3$ is fixed to three while $\cvisSQ$ and $\ceffSQ$ applied to 
one or two extra degrees of freedom are allowed to vary.
\end{table}

Since both $\cvis$ and $\ceff$ modulate the amplitude of small-scale power in the CMB we 
explored degeneracies with the running of the spectral index, $\alpha_s 
\equiv d n_s/d\ln k$.  Excluding the Lyman-alpha data and fixing 
$\cvisSQ = \ceffSQ = 1/3$ we find,  $\alpha_s = -0.020 \pm  0.013 \pm 0.026$ and $\neff = 3.53\pm 0.21\pm0.72$.  
Allowing both $\cvis$ and $\ceff$ to vary we find  $\alpha_s =-0.019\pm  0.016\pm0.03$, $\neff = 3.49 \pm 0.20^{+0.73}_{-0.70}$, 
$\cvisSQ = 0.29 ^{+0.05+0.27}_{-0.08-0.19}$and $\ceffSQ = 0.33 \pm  0.02^{+0.05}_{-0.04}$ .  

Although there are many radiative backgrounds which are unrelated to neutrinos, if the anomalous CDRB is explained by a modification to neutrino physics this may lead to a change the functional form of $Y_p(\neff)$.  Therefore it is important to consider 
the case where $Y_p$ is an additional free parameter.  
Fixing 
$\cvisSQ = \ceffSQ = 1/3$ we find $Y_p = 0.294 \pm 0.033^{+0.064}_{-0.067}$ and $\neff = 
3.64^{+ 0.21+0.86}_{- 0.24-0.79}$ which is in slight ($\sim$1.5$\sigma$) disagreement with astrophysical measurements \cite{Ypmeas} and BBN predictions for $Y_p$ \cite{Steigman:2007xt}.  However, when we also allow $\cvis$ and $\ceff$ to vary, the helium fraction 
preferred by the data decreases to $Y_p = 0.257\pm 0.051\pm0.1$  which is in good agreement with 
the most recent astrophysical measurements, $Y_p = 0.2565 \pm0.0010\ ({\rm stat}) \pm 0.0050\ ({\rm syst})$ \cite{Ypmeas}. 
We find the number of 
effective neutrino species is $\neff = 3.73^{+0.24+0.98}_{- 0.28-0.89}$ with $\cvisSQ = 0.33^{+ 0.04+0.22}_{- 0.06-0.16}$ and $\ceffSQ = 0.315 \pm 0.018^{+0.037}_{-0.033}$.  Using the astrophysical measurement of $Y_p$ as a prior we find $\neff = 3.73 \pm 0.2 \pm 0.7$, $\cvisSQ = 0.34^{+ 0.03+0.21}_{-0.06 - 0.15}$, and $\ceffSQ = 0.313 \pm 0.014 ^{+0.028}_{-0.030}$. 

We note that our modified perturbation equations only apply to massless degrees of freedom.  However, since  a non-zero mass predominately affects the late-time (post-recombination) evolution of the perturbations, its effects  
are separated in time (and hence we expect should be fairly uncorrelated) from the effects of varying $\cvis$ and $\ceff$, which are most important before and during recombination.  We leave a simultaneous constraint 
on the neutrino mass, $\cvis$ and $\ceff$ to future work. 

\section{Conclusions}

We have used current CMB and LSS data to explore various properties of 
perturbations in the CDRB.  In particular, we have parameterized the 
evolution of the neutrino perturbations with two additional parameters: a 
rest-frame sound speed, $\ceffSQ$, and a viscosity parameter $\cvisSQ$, which both 
equal 1/3 for the standard, non-interacting, CDRB.  With $\cvisSQ = \ceffSQ=1/3$ we find that 
current data favors an anomalously large standard CDRB energy density at the 99.9\% C.L.. 
When $\cvis$ and $\ceff$ are allowed to vary the data still shows that the 
standard value $(\neff,\ceffSQ) =(3,1/3)$ is disfavored at slightly greater than 
the 95\% C.L. relative to the 2D-marginalized contours.   

Our results can be interpreted as providing tentative evidence that the extra relativistic degrees of 
freedom seen in observations of the CMB 
may have non-negligible interactions with $\ceffSQ < 1/3$ at the  87.5\% C.L.  Although, as shown in Fig.~\ref{fig:constraints}, this result is driven 
by the small-scale observations of the CMB and may be impacted by systematic errors, such as 
issues related to marginalizing over sources of secondary anisotropies.  A more conclusive result 
must wait for data from future observations, such as the Planck satellite \cite{:2011ah}.  

If any anomalous radiative energy density is due to a modification of the neutrino sector this may result in a change to the standard relationship $Y_p(\neff)$.  Allowing the helium fraction to also 
vary we find that the constraint on $Y_p$ is in agreement with the most recent astrophysical measurements \cite{Ypmeas}.  In addition the values of $\ceffSQ$ and $\cvisSQ$ are consistent with their non-interacting value of 1/3 and $\neff$ is larger than the expected value of 3 at the 90\% C.L. Using the astrophysical measurement of $Y_p$ as a prior we find $\neff >3$ at the 95 \% C.L., $\cvisSQ$ is fully consistent with the expected value of 1/3, and $\ceffSQ$ is less than 1/3 at the $85\%$ C.L.  Therefore we find that although fixing the helium fraction through its BBN relationship $Y_p(\neff)$ may not be appropriate in general, when we allow $Y_p$ to be a free parameter the constraint on $\neff$ is still significantly anomalous.  Using the astrophysical measurement as a prior on $Y_p$ only increases this significance and hints at a slightly low value for $\ceffSQ$. 
 
Using a Fisher analysis we find that future observations of the CMB with the
Planck satellite alone will provide a measurement of
$\neff = 3.0 \pm 0.17$, $\cvisSQ = 0.333 \pm 0.026$, and $\ceffSQ = 0.333 \pm 0.004$.
If future observations continue to provide evidence for the presence 
of extra relativistic energy 
density then, when applied to one additional effective neutrino degree of freedom, Planck will 
constrain $\cvisSQ = 0.3 \pm 0.1$ and $\ceffSQ = 0.333 \pm 0.017$. The improved sensitivity to 
these parameters will allow a constraint on the
fundamental properties of any new radiative degrees of freedom. 

\emph{Note added in proof:} after a preprint of this paper appeared on the arXiv we became aware of a study, Ref.~\cite{Archidiacono:2011gq}, which presents similar results. 

\begin{acknowledgments}
TLS thanks Daniel Grin, Roland De Putter, and Zane Smith for useful conversations.   Some computations were performed on the GPC supercomputer at the SciNet HPC Consortium.  This research used resources of the National Energy Research
Scientific Computing Center, which is supported by the Office of
Science of the U.S. Department of Energy under Contract No.~DE-AC02-05CH11231.  This research is supported by the Berkeley Center for Cosmological Physics.  
\end{acknowledgments}

\begin{appendix}
\section{Initial conditions}
Following the derivation outlined in Ref.~\cite{ma_bert} we set the initial conditions to the growing mode which reverts 
to the standard adiabatic 
mode when $\cvisSQ =\ceffSQ = 1/3$.  For reference, these initial conditions are given in synchronous gauge by 
\begin{widetext}
\begin{eqnarray}
\delta_c &=& \delta_b =  \frac{3}{4} \delta_{\nu}^{MV} = \frac{3}{4} \delta_{\gamma} =- \frac{\chi}{3}k^2 \tau^2, \\
\delta_{\nu}^{ML} &=& \left\{ \frac{2[10+\cvisSQ (2+7 R_{\nu}^{ML} + R_{\nu}^{MV})]-3 \ceffSQ 
[5+2 \cvisSQ (2+R_{\nu}^{ML} + R_{\nu}^{MV})]}{10+3 \ceffSQ + 6 (1+3 \ceffSQ) \cvisSQ R_{\nu}^{ML}}\right\} \delta_{\gamma}, \\
q_{\gamma} &=& \frac{k \tau}{9} \delta_{\gamma}, 
\end{eqnarray}
\begin{eqnarray}
q_{\nu}^{ML} &=& \left\{ \frac{2 \cvisSQ (2 + R_{\nu}^{ML} + R_{\nu}^{MV} + 3 \ceffSQ^2[5+2 \cvisSQ
(2 +R_{\nu}^{ML} +R_{\nu}^{MV})]}{30 + 45 \ceffSQ + 18(1+3 \ceffSQ) \cvisSQ R_{\nu}^{ML}}\right\}  k\tau \delta_{\gamma} \\
q_{\nu}^{MV} &=& \frac{1}{9} \left(1 + \frac{4 (2+R_{\nu}^{ML})}{15 + 4 R_{\nu}^{MV}}\right)k \tau\delta_{\gamma}, \\
\pi_{\nu}^{ML} &=& \chi \frac{2 \cvisSQ \{\ceffSQ [6 R_{\nu}^{ML} - 3(2+R_{\nu}^{MV})]-2(2+R_{\nu}^{ML} + R_{\nu}^{MV})\}}{10+3 \ceffSQ + 6 (1+3 \ceffSQ) \cvisSQ R_{\nu}^{ML}}k^2\tau^2, \\
\pi_{\nu}^{MV} &=& \chi\frac{2(2+R_{\nu}^{ML}}{15 + 4 R_{\nu}^{MV}} k^2\tau^2, \\
z &=& -\frac{3}{2} \delta_{\gamma},
\end{eqnarray}
\end{widetext}
where $ML$ denotes the massless neutrinos (parameterized by $\ceff$ and $\cvis$), $MV$ denotes the standard massive neutrinos, $\delta_i$ are the density contrasts, $q_i$ is the heat flux, $\pi_i$ is the anisotropic stress, $z = \frac{1}{2}\dot{h}$ where $h$ is the standard synchronous 
gauge potential (see Ref.~\cite{ma_bert}), $R_{\nu}^{ML} = \rho_{\nu}^{ML}/\rho^{\rm tot}_{\rm rad}$ is the fraction of the total radiation energy density in 
massless (non-standard) neutrinos, and $R_{\nu}^{MV} = \rho_{\nu}^{MV}/\rho^{\rm tot}_{\rm rad}$ is the fraction of the total radiation energy density 
in massive, standard, neutrinos.  One can check that that for standard neutrinos ($\ceffSQ = \cvisSQ = 1/3$) these initial conditions revert back to the 
standard adiabatic initial conditions; for $\ceffSQ \neq 1/3$ the initial conditions are an admixture of adiabatic and isocurvature initial conditions. 

\end{appendix}

\end{document}